\documentclass[twocolumn,preprintnumbers,secnumarabic,amsmath,amssymb,nofootinbib,floatfix]{revtex4}
\usepackage{varioref,exscale,latexsym,amsmath,amssymb}
\usepackage{graphicx,epstopdf}

\usepackage{graphicx}
\usepackage{epstopdf}
\usepackage{dcolumn}
\usepackage{bm}

\def\beq{\begin{equation}}

\def\eeq{\end{equation}}

\def\beqa{\begin{eqnarray}}

\def\eeqa{\end{eqnarray}}

\begin{document}

\title{{\bf Photon-photon scattering and tests of gauge invariance  }}

\medskip\
{\author{Basem Mahmoud El-Menoufi}
\author{ John F. Donoghue}
\affiliation{Department of Physics,
University of Massachusetts\\
Amherst, MA  01003, USA}

\begin{abstract}
We consider the phenomenology of a dimension-four operator that violates electromagnetic gauge invariance. Its magnitude is severely constrained by the lack of scattering of very low energy electromagnetic radiation off of the Cosmic Microwave Background (CMB) and by the lack of an induced mass when photons propagate in the CMB. We also discuss possible Lorentz-violating extensions of the operator basis. If a bare Proca mass exists and dominates over the induced mass, there is also a tight constraint from high energy scattering.
\end{abstract}
\maketitle
\section{Introduction}

Given that our present theories are built using the theme of gauge invariance, it is of some interest to understand the experimental limits on gauge invariance. Moreover, there are motivations for possible small violations of gauge invariance, discussed below in Sec. 2 In this paper we study the effects of the Lagrangian
\begin{equation}
{\cal L}_{gv} = -\frac18 \kappa A_\mu A^\mu A_\nu A^\nu
\end{equation}
which represents the next gauge-violating operator after a photon mass term. We will derive constraints of order $\kappa\le 10^{-15}$ from the scattering of very low energy electromagnetic radiation, $\kappa\le 10^{-28}$ from the generation of Lorentz-violating photon mass when propagating in the Cosmic Microwave Background (CMB) and yet stronger constraints in the possible presence of a Lorentz-conserving photon mass.

The operator of Eq. 1 leads to a cross section for photon-photon scattering that diverges at low energy. This can be seen simply from the dimensionless coupling constant $\kappa$ and the absence of a mass such that the total cross section must go as
\begin{equation}
\sigma \sim \frac{\kappa^2}{s}
\end{equation}
simply on dimensional grounds. Here $s=E_{cm}^2$ is the square of the center of mass energy. Such an interaction would scatter the lowest energy electromagnetic radiation. In particular, radio waves from distant objects would scatter off the Cosmic Microwave Background (CMB) and would directly not reach us if the mean free path is too short. We study this effect in Sec 3 and use it to bound $\kappa$. Section 4 considers some Lorentz violating variants of this interaction.

Moreover, when propagating through the Cosmic Microwave Background (CMB) a photon will pick up an effective mass term from this interaction. This mass is not Lorentz invariant as it depends on the rest frame of the CMB. Somewhat remarkably, such a Lorentz-violating mass is better behaved than a Lorentz-invariant one, and has interesting phenomenology which we explore in Sec. 5. Finally we explore the possible enhancement of the limits that happens for some values of a Lorentz-conserving mass term, which occur because scattering from transverse to longitudinal gauge bosons occurs with a cross section inversely proportional to the possible photon mass.

\section{Brief comments on gauge non-invariance}

In looking for signals of new physics, studies of violations of symmetries are particularly useful. Much of the focus of searches for physics beyond the standard model uses effective Lagrangians of dimension greater than four, which preserve the gauge symmetries of the Standard Model, but allow the violation of any global symmetry\cite{Buchmuller:1985jz}. Another significant sub-field looks for violations of Lorentz invariance. This involves a further extension of the Standard Model\cite{Colladay:1998fq}, this time involving Lorentz violating operators of dimension 2, 3, 4 etc, still with the assumption that the Standard Model gauge symmetries are exact. Here we are concerned with a piece of phenomenology giving up the last of the symmetries - gauge symmetry.

One motivation comes from the connection between Lorentz invariance and gauge invariance. It is a little appreciated feature that the massless gauge fields are not themselves Lorentz covariant\cite{WeinbergQFT}. The Lorentz transformation needs to be supplemented by a gauge transformation in order to make an overall invariance. A simple example is a Lorentz boost of a transverse photon polarization vector $\epsilon_\mu (k)$ in a direction that is not along the initial propagation direction $k_\nu$, resulting in a polarization vector which is no longer transverse to the propagation direction in the new frame. A gauge transformation is required to make this polarization vector transverse.  Because of this connection it is possible that loop effects with Lorentz violating interactions could produce gauge non-invariant effects. Indeed such effects are found in specific calculations\cite{Altschul:2003ce}-\cite{Jackiw:1999qq}. There are some controversies concerning the regularizat
 ion of the gauge-violating effects\cite{Jackiw:1999qq} but a reasonable conclusion could be that the generation of such effects is possible depending on the nature of the ultimate high energy theory.

Another motivation is the possibility of emergent symmetries\cite{emergent}. There are known examples in condensed matter systems where a gauge symmetry emerges in the ground state of a theory that does not originally have such a symmetry. There are also many attempts at emergent space-time, which would require that the general covariance of gravity be emergent also. In such cases, the leading approximation to the full theory would satisfy the emergent gauge symmetry. However, since the underlying theory does not respect the symmetry, there can be small corrections to the leading approximation violating the symmetry. At the very least, loop diagrams should be able to probe the lack of symmetry at the fundamental scale, as the loops involve integrations over all energies. Looking for violations of gauge invariance is the part of the phenomenology of emergent gauge symmetry\cite{Donoghue:2010qf}.

The Weinberg-Witten theorem \cite{Weinberg:1980kq, Loebbert:2008zz} connects these two motivations as it says that it is not possible to have a composite or emergent gauge boson\footnote{The theorem applies to Yang-Mills gauge bosons but not to those of a U(1) symmetry. However, in the Standard Model the physical photon is a mixture of SU(2) and U(1) fields so that we expect that the theorem is applicable to real photons.} or graviton from a Lorentz invariant theory. This implies that emergent gauge symmetry would come from a Lorentz violating theory, and Lorentz symmetry itself could be emergent.

The theorem does not give any indication of how large any residual violations are, but it make sense to search for gauge symmetry violation and Lorentz violation together. Because of this connection we will also briefly study a generalization of Eq. 1,
\begin{equation}
\mathcal L=-\frac{1}{4}F^2-\frac{1}{8}\kappa_{\mu\nu\alpha\beta}A^{\mu}A^{\nu}A^{\alpha}A^{\beta}
\end{equation}
with a more general tensor $\kappa_{\mu\nu\alpha\beta}$. Eq. 1 is reproduced with
\begin{equation}
\kappa_{\mu\nu\alpha\beta}=\kappa g_{\mu\nu} g_{\alpha\beta}  ~~.
\end{equation}
and we will consider the case where there is a preferred direction in space $b_\mu$, with the possibilities
\begin{align}
\kappa_{\mu\nu\alpha\beta}=\kappa' b_{\mu} b_{\nu} b_{\alpha} b_{\beta}
\end{align}
and
\begin{equation}
\kappa_{\mu\nu\alpha\beta}=\kappa'' b_\mu b_\nu g_{\alpha\beta}  ~~.
\end{equation}

\section{Scattering off the Cosmic Microwave Background}

Photons from a distant source can scatter off of the CMB. With the usual gauge invariant interaction, the Euler Heisenberg Lagrangian, this becomes significant only at the highest energies, leading the the well-known GZK cutoff of ultra-high-energy cosmic rays\cite{Greisen:1966jv}. However, with the interaction of Eq. 1, the strongest scattering is at low energy.

Let us calculate the cross-section for $\gamma\gamma\rightarrow\gamma\gamma$ scattering using Eq. 1.  The matrix element is found to be
\begin{align}
\nonumber
\mathcal M=\kappa &[\epsilon(\boldsymbol{p}_1,\lambda_1)\cdot \epsilon(\boldsymbol{p}_2,\lambda_2)\epsilon(\boldsymbol{p}_3,\lambda_3)\cdot \epsilon(\boldsymbol{p}_4,\lambda_4)\\\nonumber
+&\epsilon(\boldsymbol{p}_1,\lambda_1)\cdot \epsilon(\boldsymbol{p}_3,\lambda_3)\epsilon(\boldsymbol{p}_2,\lambda_2)\cdot \epsilon(\boldsymbol{p}_4,\lambda_4)\\
+&\epsilon(\boldsymbol{p}_1,\lambda_1)\cdot \epsilon(\boldsymbol{p}_4,\lambda_4)\epsilon(\boldsymbol{p}_2,\lambda_2)\cdot\epsilon(\boldsymbol{p}_3,\lambda_3)]
\end{align}
At this stage we do not consider a photon mass term, returning to this topic in later sections. Without a mass, we consider only the production and scattering of transverse photons. Operationally, this means the polarization vectors are 'physical',
\begin{align}
\epsilon^0(\boldsymbol{p},\pm1)&=0\\
\vec{p}\cdot \vec{\epsilon}(\boldsymbol{p},\pm1)&=0
\end{align}
Because the interaction is not gauge invariant, the usual replacement $\sum_{polarization} \epsilon_{\mu} \epsilon_{\nu} \rightarrow -g_{\mu\nu}$ can not be made here. This simple replacement only works in QED because of gauge-invariance leading to the cancellation of the contributions coming from the unphysical longitudinal and scalar degrees of freedom. To square the invariant amplitude then, we directly perform the calculation in the CM frame for each polarization configuration individually. To this end, we employ a linear polarization basis with the following convention,
\begin{align}
\vec{\epsilon}(\boldsymbol{p},1) \times \vec{\epsilon} (\boldsymbol{p},2) = \hat{p}
\end{align}
The non-vanishing transitions are shown in Table I.
\begin{table}[h]
\caption{Polarizations and Matrix Elements}
\centering
\begin{tabular}{c c}
\hline
Config. & $\mathcal M$ \\
\hline
$12\rightarrow11$ & $\kappa \cos \theta_{cm}$ \\
$22\rightarrow22$ & $\kappa \cos \theta_{cm}$ \\
$11\rightarrow22$ & $\kappa \cos \theta_{cm}$ \\
$22\rightarrow11$ & $\kappa \cos \theta_{cm}$ \\
$21\rightarrow21$ & $3\kappa$ \\
$21\rightarrow12$ & $\kappa$ \\
$12\rightarrow21$ & $\kappa$ \\
$12\rightarrow12$ & $\kappa \left(1+2\cos^2 \theta_{cm} \right)$ \\
\hline
\end{tabular}
\label{tab:ME}
\end{table}
We sum (average) over the initial (final) polarization states and obtain the modulus-squared amplitude,
\begin{align}
\frac{1}{4}\sum_{polarizaton}|\mathcal M|^2=\kappa^2(3+2\cos^2\theta_{cm}+\cos^4\theta_{cm})
\end{align}
where $\theta_{cm}$ is the angle the outgoing particles makes with the collision axis. The differential scattering cross-section is readily found to be,
\begin{align}
\frac{d\sigma}{d\Omega_{cm}}=\frac{\kappa^2}{64 \pi^2 s}(3+2\cos^2\theta_{cm}+\cos^4\theta_{cm})
\end{align}
which upon integration yields the total cross-section,
\begin{align}
\sigma=\frac{29 \kappa^2}{120 \pi s}
\end{align}

In order to obtain a bound on the coupling constant $\kappa$, we consider the scattering of a low energy photon ($E_{\gamma}$) emmitted from a distant quasar off a CMB photon ($E$) and calculate the collision rate for such a reaction. Obviously we perform the calculation in the earth frame since all energy measurements are known only with respect to the aforementioned frame. We align the momentum of the incoming photon along the z-axis and consider the momentum of the CMB photon to make an angle $\theta$ with the z-axis. Thus we have,
\begin{align}
\nonumber
s&=2EE_{\gamma}(1-\cos \theta)\\
&=4EE_{\gamma}\sin^2{\frac{\theta}{2}}
\end{align}
The CMB photons can impinge on the incoming photon at any angle and any energy. Moreover, they obey Bose-Einstein statistics and we average appropriately over all angles and energies. The total cross-section diverges because of the pole at $\theta =0^\circ$. This comes from the limit of collinear photons. However, we will see that the total collision rate is appropriately finite.

The total collision rate, whose inverse yields the mean free path, can be used to bound $\kappa$. Following \cite{Liang:2011sj}, we define the number density of CMB photons as
\begin{align}
&d\rho(E,\theta,\phi)=\frac{E^2}{4\pi^3}\frac{dE d\phi d\theta \sin\theta}{\left(\exp{\beta E}-1\right)}
\end{align}
The rate of collision between the incoming photon and CMB photons becomes
\begin{align}
d\Gamma=|\vec{v}_\gamma-\vec{v}_{E}|\sigma d\rho(E,\theta,\phi)
\end{align}
where $|\vec{v}_{\gamma}-\vec{v}_{E}|$ is the relative speed in the earth frame.
\begin{align}
\vec{v}_{\gamma}-\vec{v}_{E}=-& \left(\sin \theta \cos \phi,\sin \theta \sin \phi,\cos \theta-1\right)\\
& |\vec{v}_{\gamma}-\vec{v}_{E}|=2\sin{\frac{\theta}{2}}
\end{align}
Finally we integrate over all angle and energies to obtain the total collision rate,
\begin{align}
\Gamma=\frac{29\kappa^2}{120\pi^3}&\int dE d\cos{\theta}\frac{E^2\sin{\frac{\theta}{2}}}{4EE_{\gamma}\sin^2{\frac{\theta}{2}}\left(e^{\beta E}-1\right)}\\
&\Gamma=\frac{29\kappa^2\zeta(2)}{120\pi^3E_{\gamma}\beta^2}
\end{align}
Here
\begin{align}
\zeta(2)=1.645\\
\end{align}

It is well known that very low-energy radio signals reach the earth coming from distant quasars, which are billions of light-years away from us. We place an upper bound on $\kappa$ by demanding that the time taken by these low-energy signals to reach us has to be less than the time between individual collisions, given by the inverse of the collision rate.

The most distance known radio-loud quasar is the source J1427+3312 \cite{Frey:2008ix}. It has been detected at 1.6 GHz with a redshift of z=6.12. The light-travel distance is simply calculated from the redshift, and thus the light-travel time is
\begin{align}
t \approx 4 \times 10^{17} sec
\end{align}
We also work with the average,
\begin{align}
\langle \beta^2 \rangle = \frac{k^{-2}_B}{(2.725)^2(z+1)} \approx 2.547 \times 10^6 (eV)^{-2}
\end{align}
In the above, we used $T_{CMB}=2.725(1+z)$. This leads to the bound
\begin{align}
\kappa \leq 3.686 \times 10^{-15}  ~~~.
\end{align}
In fact we can do slightly better by using a closer source which is however observed at a lower frequency. For example in reference \cite{Shang:2011qm}, the sources in the sample had observations in the frequency as low as 74 MHz. The highest red-shift among the radio-loud quasars is the source 3C 446 with z=1.404. We can use the data of this source to find
\begin{align}
T \approx 2.857 \times 10^{17} sec \quad , \quad \langle \beta^2 \rangle \approx 7.556 \times 10^6 (eV)^{-2}
\end{align}
Then the final bound is
\begin{align}
\kappa \leq 1.615 \times 10^{-15}~~~.
\end{align}
We see that the trade-off between energy and red-shift produces an $\mathcal{O} (1)$ effect. Finally, we note that it is very likely that space-based observations could detect quasars in lower regimes of the radio spectrum which greatly enhances our bound.

\section{Combined Lorentz and gauge violation}

Because of the possible connection of Lorentz and gauge violation, we here consider the combined violation of both types of symmetries. The resulting bounds are not much different, but there are some interesting features which emerge when both symmetries are broken.

Lorentz violation is generally parameterized by a fixed background vector (or tensor) that specifies a preferred direction\cite{Colladay:1998fq}. For our studies we use a particular preferred frame designated by $b_{\mu}$. In this case the preferred frame can also enter the gauge violating interaction.

We first consider the following coupling,
\begin{align}
\kappa_{\mu\nu\alpha\beta}=\kappa' b_{\mu} b_{\nu} b_{\alpha} b_{\beta} \ \ .
\end{align}
If we again consider the scattering of transverse photons with this modified coupling, we find for the matrix element
\begin{align}
\mathcal{M}=3 \kappa' \left[ b \cdot \epsilon(\vec{p}_1,\lambda_1) \quad b \cdot \epsilon(\vec{p}_2,\lambda_2) \quad b \cdot \epsilon(\vec{p}_3,\lambda_3) \quad b \cdot \epsilon(\vec{p}_4,\lambda_4)\right] \ \ .
\end{align}

To compute the cross section, we carry out the polarization sum in the CM frame. To this end, we use the following polarization sum formula \cite{WeinbergQFT}
\begin{align}
\sum_{\lambda=1}^{2} \epsilon^i(\vec{p},\lambda) \epsilon^j(\vec{p},\lambda)=\delta^{ij}-\frac{p^i p^j}{\vec{p}^2}
\end{align}
The squared matrix element is then,
\begin{align}
\mathcal{M}^2=\frac{9}{4}\kappa'^2 \left(\vec{b}^2-b^2_3\right)^2 \left(\vec{b}^2-(\vec{b} \cdot \hat{p}_3)^2\right)^2
\end{align}
Where $b_3$ is the component of $\vec{b}$ along the collision axis in the CM frame. Notice that the differential cross section is not azimuthally symmetric in this case as the background vector field defines a preferred direction in any single frame. The cross section is then,
\begin{align}
\sigma=\frac{3 \kappa'^2 \vec{b}^4}{80 \pi s}\left(\vec{b}^2-b^2_3\right)^2
\end{align}

Let us for completeness also consider the following coupling,
\begin{align}
\kappa_{\mu\nu\alpha\beta}=\kappa'' g_{\mu\nu}b_{\alpha}b_{\beta}
\end{align}
As before, $\kappa''$ is the parameter controlling the strength of gauge-invariance violation, while $b_{\mu}$ is a constant four-vector field controlling Lorentz-invariance violation.

We follow the same procedure as last section and compute the matrix element for each polarization configuration individually in the CM frame. As the matrix element for all polarization configurations are non-trivial, we choose to compute the cross section for a specific configuration, namely $\mathcal{M}(11 \rightarrow 11)$ where we again used a linear polarization basis with the conventions of the last section.
\begin{eqnarray}
\sigma(11 \rightarrow 11)&=&\frac{\kappa''^2}{15360 \pi s} \left(509 b^4_1 + 9 b^4_2 + 64 b^4_3 \right. \nonumber \\
&~&~\left.+142 b^2_1 b^2_2 + 432 b^2_1 b^2_3 + 18 b^2_2 b^2_3 \right)
\end{eqnarray}

The inclusion of Lorentz violation does not generate any improvement in the bounds. The constraints from the mean free path go through as before. We neglect the potential dependence on the relative direction of a particular quasar to the spatial direction of $\vec{b}$ and then, taking into account the differing numerical factors in the cross sections, quote rough bounds
\begin{equation}
\kappa' b^4 < 10^{-14}~, ~~~~~ \kappa'' b^2 < 10^{-14}  \ \ .
\end{equation}

We note that the combined Lorentz and gauge violation posesses a very interesting feature, namely that it is not invariant under the so-called `observer' Lorentz transformations which is just the conventional class of transformations of special relativity. As discussed before, the polarization vectors of the gauge-field are not proper four vectors, thus the scalar products in the matrix element are not strictly Lorentz invariant. This is different than the framework of the Lorentz violating SME, which is invariant under the aforementioned class of transformations but not invariant under the so called `particle' Lorentz transformations, where breaking the symmetry comes about by the existence of background constant tensor fields which define a preferred direction in any single frame. As shown, adding gauge-violation to the framework of SME adds this new feature which enables us to probe Lorentz-symmetry violation in the conventional sense. A final remark worth mentioning is th
 at the pure gauge-violating theory does not exhibit such a feature because the matrix element includes only scalar products of polarization vectors. In other words, the field is purely self-interacting so that Lorentz-invariance violation is `hidden' but in the present case, the background vector acts like an external field coupled to the gauge-field and thus the result we have just shown.\\

\section{Mass generation from the CMB}

Once one considers a gauge non-invariant interaction such as Eq. 1, there is no symmetry forbidding a photon mass. However, the interaction itself does not have to generate a photon mass in a vacuum. A loop diagram involving two of the photons of the 4 photon vertex would have the potential to generate a mass term. However, the loop integral vanishes when regularized dimensionally. Despite this, the interaction of Eq. 1 will {\em} necessarily generate a mass for a photon moving in a background of other photons, i.e. the CMB. We evaluate this mass in this section and explore some of its more unusual properties.

We treat the heatbath as a background field which the photon is propagating in. This method is equivalent to a photon loop in real-time finite temperature field theory. We treat the interaction Lagrangian by splitting the field into a background piece - the heat bath - and a quantum propagating photon,
\begin{equation}
A^{\mu}=A_b^{\mu}+A_Q^{\mu}
\end{equation}
Expanding the product of fields in the Lagrangian, we collect the six terms quadratic in the background field.
\begin{align}
\nonumber
\mathcal{L}^{quad}=-\frac{\kappa_{\mu\nu\alpha\beta}}{8}(&A_b^{\mu}A_b^{\nu}A_Q^{\alpha}A_Q^{\beta}+A_b^{\mu}A_b^{\alpha}A_Q^{\nu}A_Q^{\beta}\\\nonumber
+&A_b^{\mu}A_b^{\beta}A_Q^{\alpha}A_Q^{\nu}+A_b^{\alpha}A_b^{\nu}A_Q^{\mu}A_Q^{\beta}\\\nonumber
+&A_b^{\beta}A_b^{\nu}A_Q^{\alpha}A_Q^{\mu}+A_b^{\alpha}A_b^{\beta}A_Q^{\mu}A_Q^{\nu})
\end{align}
The finite-tempreture ground state is defined as,
\begin{align}
\langle \beta |a^{\dagger}(\vec{p},\lambda)a(\vec{q},\lambda^{\prime})| \beta \rangle=n_B (E_{\vec{p}}) (2\pi)^3 \delta_{\lambda \lambda^{\prime}}\delta^3(\vec{p}-\vec{q})
\end{align}
Where $n_B$ is the usual Bose-Einstein distribution for photons. The contribution of the CMB would be calculated by the taking the ground state expectation value of the quadratic Lagrangian. First, we evaluate
\begin{equation}
\langle \beta | A_b^{\mu}A_b^{\nu} | \beta \rangle=\sum_{\lambda}\int \frac{d^3p}{(2\pi)^3 E_p}n_B(E_p)\left[\epsilon^{\mu}(\boldsymbol{p},\lambda)\epsilon^{\nu}(\boldsymbol{p},\lambda)\right]
\end{equation}
To evaluate the above integral, we note that from symmetry the tensor integral is proportional to the Kronecker delta.
\begin{align}
\sum_{\lambda}\int \frac{d^3p}{(2\pi)^3 E_p}n_B(E_p)&\left[\epsilon^{i}(\boldsymbol{p},\lambda)\epsilon^{j}(\boldsymbol{p},\lambda)\right]=a\delta^{ij}
\end{align}
Contracting both sides with $\delta_{ij}$, we find
\begin{align}
&a=\frac{2}{3}I
\end{align}
Where $I$ is a simple integral over the BE distribution,
\begin{equation}
I =\int \frac{d^3p}{(2\pi)^3 E_p}n_B(E_p) =\frac{\zeta(2)}{2\pi^2}\beta^{-2}
\end{equation}
Thus the quadratic Lagrangian reads,
\begin{align}
\mathcal{L}^{quad}=\frac12\kappa I \left(A^2_Q-\frac{2}{3}\vec{A}_Q\cdot \vec{A}_Q\right)
\end{align}
Now we construct the free field Lagrangian for the quantum field, which reads
\begin{align}
\mathcal{L}_0=-\frac{1}{4}F^2+\frac{1}{2} m_{\alpha\beta}A^\alpha A^\beta
\end{align}
Where we defined,
\begin{align}
m_{00}\equiv \kappa I \quad m_{ij}\equiv -\frac{5\kappa I}{3}\delta_{ij}
\end{align}
We see that $\kappa$ has to be positive definite to ensure the theory does not have growing exponential solutions.

In order to solve the equations of motion, we consider
\begin{align}
\partial_{\mu}&F^{\mu\nu}+m_{\alpha}^{\nu} A^{\alpha}=0\\
&m_{\alpha}^{\nu}\equiv g^{\nu\beta}m_{\alpha\beta}
\end{align}
By acting $\partial_{\nu}$ on the equation of motion and requiring the external current coupled to the field to be conserved, we obtain the following constraint which ensures the field has only three degrees of freedom.
\begin{align}
\partial_{\nu}\partial_{\mu}&F^{\mu\nu}+m^{\nu}_{\alpha}(\partial_{\nu}A^{\alpha})=0\\
&m^{\nu}_{\alpha}(\partial_{\nu}A^{\alpha})=0\\
&\partial_0A^0+\frac{5}{3}\partial_iA^i=0
\end{align}
We propose the usual wave ansatz,
\begin{align}
A(x) \propto e^{-i\left(\omega t-\vec{k} \cdot \vec{x}\right)}
\end{align}
Thus Eq. 49 becomes,
\begin{align}
\frac{3}{5}\omega A^0-\vec{k}\cdot \vec{A}=0
\end{align}
Due to the  manifest Lorentz non-invaraince, the quantization procedure needs some care. First, we explore the dispersion relations by plugging the wave ansatz in the equations of motion.
\begin{equation}
\left(-\omega^2+\vec{k}^2\right)A^{\nu}+k^{\nu}(k\cdot A)+m_{\alpha}^{\nu} A^{\alpha}=0
\end{equation}
For definiteness, we take $\vec{k}=k\hat{z}$. Taking $\nu=0$ and using the constraint equation, we obtain
\begin{align}
\left(-\omega^2+\vec{k}^2\right)A^0+\omega\left(\omega A^0-\frac{3}{5}\omega A^0\right)+m_0^0A^0=0
\end{align}
This yields the following dispersion relation
\begin{equation}
\frac{3}{5}\omega^2-\vec{k}^2=\kappa I
\end{equation}
The same dispersion relation is readily obtained if $\nu=3$.\\
On the other hand, for $\nu=1,2$ we get
\begin{align}
\left(-\omega^2+\vec{k}^2 \right) A^{1,2}+\frac{5}{3}\kappa I A^{1,2}=0
\end{align}
which yields the dispersion relation,
\begin{align}
\omega^2-\vec{k}^2=\frac{5}{3}\kappa I
\end{align}
The differing dispersion relations are a clear manifestation of the Lorentz non-invariance of the theory. We next show how to quantize the theory by introducing different dispersion relations for transverse and longitudinal modes.

For later convenience, we define
\begin{align}
\frac{5}{3}\kappa I \equiv m^2
\end{align}
The aforementioned theory describes three degrees of freedom. Guided by the previous analysis , we proceed by proposing two different wave ansatz for the different modes of propagation.
\begin{align}
A_k(x)\propto \displaystyle\sum\limits_{\lambda=1}^2 e^{-i\left(\omega t-\vec{k} \cdot \vec{x}\right)}\epsilon(\lambda,\boldsymbol{k})+e^{-i\left(\omega_L t-\vec{k} \cdot \vec{x}\right)}\epsilon(3,\boldsymbol{k})
\end{align}
where four-vectors are displayed without arrows. To satisfy the equations of motion and the constraint Eq. 51, the polarization vectors have to satisfy
\begin{align}
\epsilon^0(1,\boldsymbol{k})=\epsilon^0(2,\boldsymbol{k})=0\\
\vec{\epsilon}(1,\boldsymbol{k})\cdot \vec{k}=\vec{\epsilon}(2,\boldsymbol{k})\cdot \vec{k}=0\\
\epsilon^{\mu}(3,\boldsymbol{k})=\left(1,\frac{3\omega_L}{5\vec{k}^2}\vec{k}\right)
\end{align}
The third polarization vector shows a very interesting feature; it becomes null-like at some value of the 3-momentum and hence its normalization cannot be fixed. This is another manifestation of breaking Lorentz symmetry.\\
The two dispersion relations are,
\begin{align}
\omega^2-\vec{k}^2&=m^2\\
\frac{3}{5}\omega^2_L-\vec{k}^2&=\frac{3}{5}m^2
\end{align}
The fact that Eq. 62 has the canonical form of energy-momentum relation and that it is associated with the transerve modes is tempting to interpret $m$ as the 'physical' mass.

We can employ this generated mass to set an upper bound on $\kappa$ by using the most reliable experimental upper bound on the photon mass. We use the CMB current temperature to compute $I$,
\begin{equation}
I(T_{CMB}=2.71)=45\times 10^{-10}(eV)^2
\end{equation}
The photon mass is bounded to be\cite{Particle Data Group}
\begin{equation}
m_{ph}^2\leq 1\times 10^{-36} (eV)^2
\end{equation}
leading to a bound on $\kappa$
\begin{align}
\kappa \leq 1.3 \times 10^{-28}~~~.
\end{align}

\section{Mass effects in scattering amplitudes}
\subsection{Lorentz invariant mass}

In this section we explore a possible scenario which greatly enhances our bound. The logic is that if gauge-violation is admissible we do not have any theoretical argument upon which to set the mass of the photon to zero. Rather we should use the massive Proca theory, treat the photon mass as a parameter and assign its value from experiment. In this case, there is a longitudinal polarization and we can again consider scattering from transverse to longitudinal polarizations. In this situation however, the cross-section goes as $m^{-4}$ and becomes extremely strong for low masses. For many values of the mass, this leads to a very tight constraint on $\kappa$.

The Lagrangian is,
\begin{align}
\mathcal{L}=-\frac{1}{4}F^2+\frac{m^2}{2}A^2-\frac{\kappa}{8}A^4
\end{align}
\noindent Much like Section 3, we look at the process where now massive photons traveling extra-galactic distances scatter off CMB photons. A few remarks are worth mentioning at this stage regarding our choice of considering transverse photons in the initial state. Massive vector bosons are spin-1 particles having three degrees of freedom represented by polarization four-vectors satisfying the following relations
\begin{align}
&\epsilon(\boldsymbol{k},\lambda)\cdot k=0 \quad,\quad  \epsilon(\boldsymbol{k},\lambda)\cdot \epsilon(\boldsymbol{k},\lambda^{\prime})=-\delta_{\lambda \lambda^{\prime}}\\
&\sum_{spin}\epsilon_{\mu}(\boldsymbol{k},\lambda)\epsilon_{\nu}(\boldsymbol{k},\lambda)=-g_{\mu\nu}+\frac{k_{\mu}k_{\nu}}{m^2}
\end{align}
In any frame of reference other than the rest frame, the polarization vectors can be conveniently chosen as follows; two are taken to be purely spatial and transverse to the 3-momentum following a right-hand rule. If we work with states corresponding to circular polarization, then the transverse vectors correspond to the z-projection of the spin being $\pm1$.
\begin{align}
\vec{\epsilon}(\boldsymbol{k},1)\cdot \vec{k}=\vec{\epsilon}(\boldsymbol{k},2)\cdot \vec{k}=0\\
\vec{\epsilon}(\boldsymbol{k},1)\times \vec{\epsilon}(\boldsymbol{k},2)=\hat{k}
\end{align}
While the third vector is longitudinal to the 3-momentum, which corresponds to the z-projection of the spin being $0$.
\begin{align}
\epsilon^{\mu}(\boldsymbol{k},3)=\frac{1}{m}\left(|\vec{k}|,\omega\hat{k}\right)
\label{longitudinal}
\end{align}
In a frame where $\omega \gg m$, we can expand the components of the longitudinal vector and obtain
\begin{align}
\epsilon^{\mu}(\boldsymbol{k},3)=\frac{k^{\mu}}{m}+\mathcal O \left(\frac{m}{\omega}\right)
\label{helimit}
\end{align}
We imagine those massive photons being produced at extra-galactic objects via coupling to a conserved electromagnetic cuurent and thus, the Ward identity holds. Therefore, given the expansion of the longitudinal polarization vector at high energies, any matrix element with an on-shell longitudinal photon vanishes to leading order in the energy. In an arbitrary frame where $\omega \gg m$, the polarization of a photon with four-momentum $k$ is generally a superposition of the transverse and longitudinal polarizations, but based on the last discussion the contribution from the longitudinal mode vanishes.

Following the previous comments, we calculate the cross-section for the scattering of two massive transversally polarized photons in the earth frame. For simplicity, we take the initial state to be a head-on collision aligned on the z-axis and, without loss of generality, we work with linear polarization basis. Accordingly, the initial state polarization vectors are
\begin{align}
\epsilon^{\mu}(\vec{p}_1,1)=(0,1,0,0) \nonumber \\
\epsilon^{\mu}(\vec{p}_2,1)=(0,0,1,0)
\end{align}
The fact that the collision is head-on simplifies the computation considerably, the CM frame is related to the earth frame via a simple boost along the z-axis with the boost factor found to be,
\begin{align}
\beta=\frac{\omega_1-\omega_2}{\sqrt{\omega^2_1-m^2}+\sqrt{\omega^2_2-m^2}}
\end{align}
This boost obviously does not affect the polarization vectors of the incoming photons and moreover we average over the final state polarizations using the spin sum formula Eq. 69. The squared matrix element is found to be,
\begin{align}
\frac{1}{9}&\sum_{pol}\mathcal{M}^2=\frac{\kappa^2}{9m^4}[\left(m^2+(p^1_3)^2\right)\left(m^2+(p^2_4)^2\right)\\\nonumber
&+\left(m^2+(p^2_3)^2\right)\left(m^2+(p^1_4)^2\right)+2 p^1_3p^2_3p^1_4p^2_4]
\end{align}
All momenta are measured in the CM frame.\\
We consider the high-energy limit of the reaction, and therefore we neglect the masses compared to the momenta. The high energy behavior is dominated by the production of longitudinal photons in the final state. The cross-section is then found to be,
\begin{align}
\nonumber
\sigma=&\frac{1}{4|\vec{v}_1|s}\int \frac{d^3p_3}{(2\pi)^3 2\omega_3}\frac{d^3p_4}{(2\pi)^3 2\omega_4}\\ &(2\pi)^4 \delta(p_1+p_2-p_3-p_4) |\mathcal{M}|^2
\end{align}
We carry out the integral over $d^3p_4$ to find,
\begin{align}
\sigma=\frac{\kappa^2}{1152 \pi^2 |\vec{v}_1|m^4 s} \int \frac{d^3p_3}{\omega^2_3}\delta(\omega_3-\omega_1)\left((p^1_3)^2+(p^2_3)^2\right)^2
\end{align}
Performing the phase space integral and expressing the result in terms of the Mandelstam variable, we get
\begin{align}
\sigma=\frac{\kappa^2}{2160 \pi m^4}\frac{(s-4m^2)^2}{s}  \ \ .
\end{align}
We note that if we had taken longitudinal photons in the initial state, the resulting cross section would go as $m^{-8}$. However, we do not use this in producing constraints, because we do not expect to have longitudinal photons produced in quasar emission, and would not miss them if they scattered before reaching us.

We place a bound on $\kappa$ by demanding the mean free path between collisions is greater than the distance traveled by the photon from the source to earth. Looking at the cross-section, the tightest bound is obtained by considering the scattering of very energetic photons, namely GRB signal. We use $10^5 eV$, which is the typical minimum energy in GRB's, and work with the mean CMB energy $\omega_{CMB}=6.34 \times 10^{-4}eV$. The mean free path is given by\cite{Greisen:1966jv}
\begin{align}
\lambda=(n_g \sigma)^{-1}
\end{align}
Here, $n_g$ is the mean number density of CMB photons. Although we only considered head-on collisions, we use $n_g=410.4 cm^{-3}$. The effect of this approximation is obviously of $\mathcal{O}(1)$ as the GRB energy could be well above $10^5 eV$. The typical distance of GRB sources is billions of light years, $L \sim 10^{27} cm$. Finally, we use the upper experimental bound on the Lorentz-invariant photon mass and find
\begin{align}
\kappa \leq 0.67 \times 10^{-46} ~~~.
\end{align}
As discussed in the next subsection and in the summary, this bound is only realized if the Proca mass is dominant over the Lorentz violating mass that comes from the interactions with the CMB.

\subsection{Lorentz-Violating Mass}
We now also compute the cross section of the same process but with the Lorentz-violating massive theory described in Sec 5. We have just seen that if the Proca mass is dominant, there is a bad high energy behavior from the scattering into the longitudinal polarization state. This comes from the $m^{-1}$ factor seen in  Eqs. \ref{longitudinal} and \ref{helimit}. For a Lorentz violating mass, this feature does not occur, with the inverse mass being replaced by an inverse momentum.

The initial state polarization vectors are,
\begin{align}
\epsilon^{\mu}(\vec{p}_1,1)=(0,1,0,0)\\
\epsilon^{\mu}(\vec{p}_2,1)=(0,0,1,0)
\end{align}
The computation is carried in the CMB frame since this is the preferred frame in which the theory is defined. To this end, the final state photons are longitudinally polarized,
\begin{align}
\epsilon^{\mu}(\vec{p}_3,3)=\left(1,\frac{3\omega_{L,3}}{5\vec{p}^2_3}\vec{p}_3\right)\\
\epsilon^{\mu}(\vec{p}_4,3)=\left(1,\frac{3\omega_{L,4}}{5\vec{p}^2_4}\vec{p}_4\right)
\end{align}
If the initial state is a head-on collision, then the matrix element is found to be
\begin{align}
\mathcal{M}=\frac{9\kappa \omega_{L,3} \omega_{L,4}}{25 \vec{p}^2_3 \vec{p}^2_4}\left(p^1_3 p^2_4+p^2_3 p^1_4\right) \ \ .
\end{align}
To simplify the evaluation of the phase space integral, we proceed by taking $\omega_1=\omega_2$ exploiting the freedom to choose the energy of the interacting photons, since the universe is filled with photons propagating at arbitrary energies over all the electromagnetic spectrum.\\
\\
The squared matrix element is then,
\begin{align}
\mathcal{M}^2=\frac{324 \kappa^2 \omega^4_{L,3}}{625 \vec{p}^4_4}\sin^4\theta_3 \sin^2\phi_3 \cos^2\phi_3
\end{align}
Peforming the phase space integral to find the total cross section,
\begin{align}
\sigma=\frac{9 \sqrt{3} \kappa^2}{4000 \pi} \frac{s}{\left(s-4m^2\right)^2}
\end{align}

In contrast to the Proca case, this cross section is well behaved at high energy. The Lorentz violating nature of the third polarization state softens the high energy behavior. We therefore do not obtain any extra constraint from high energy scattering if the Lorentz violating mass is dominant over the Proca mass.

\section{Summary of bounds}

\begin{figure}[ht]
\begin{center}
\includegraphics[width=0.5\textwidth]{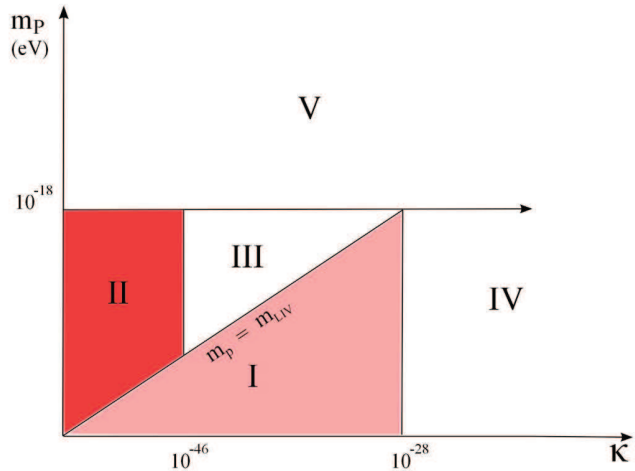}
\end{center}
\caption{Domain of $\kappa$ and the Proca mass $m_p$. As explained in the text, regions I and II are allowed and regions III, IV, and V are forbidden}
\label{D}
\end{figure}

Figure~\ref{D} shows the allowed domain of $\kappa$, which is entangled with the relative size of the possible Proca mass. Let us explain the constraints given here. Regions IV and V are excluded because the photon mass is too large. For region V, this is the standard exclusion region of a Proca mass. Region IV corresponds to the constraint from generating a Lorentz violating mass through the interactions with the CMB, as described in Sec 5. For combinations of $\kappa$ and the Proca mass that satisfy the above constraint, the situation depends on the relative size of the Proca mass versus the Lorentz violating mass from interactions with the CMB. If the Proca mass is dominant, there is the constraint from high energy scattering into the longitudinal polarization state, described in Sec 6.1, that rules out region III. However, if the Lorentz violating mass is dominant, then the longitudinal polarization state is well behaved at high energy and there is no such constraint. This yields region I as an allowed region. Finally in region II, $\kappa$ is so small that it satisfies all constraints.

The interactions of the photons thus provide extremely strong constraints on the dimensionless parameter $\kappa$ that governs the strength of this gauge violating interaction. This of course is consistent with the standard expectation that such an operator is forbidden by gauge invariance. Combined with constraints on the photon mass, our work helps quantify the degree that this expectation is supported by evidence. Of course, it is possible that gauge violation might only show up in higher dimensional operators. The mass term carries dimension two and our operator is dimension four, both of which are the dimensions allowed for renormalization interactions. If the gauge symmetry is emergent, it is plausible that at dimension two and four one obtains exactly the gauge invariant theory and that the signal for gauge violation will be higher dimensional operators suppressed by powers of some heavy scale. Such operators are also worthy of study, although the constraints are expected to be significantly weaker.

\section*{Acknowledgments} This work was partially supported by a grant from the Foundational Questions Institute and by the National Science Foundation grant PHY-0855119. Basem Mahmoud El-Menoufi would also like to thank Hansung Gim for discussions.


\begin{thebibliography}{99}


\bibitem{Buchmuller:1985jz}
  W.~Buchmuller and D.~Wyler,
  ``Effective Lagrangian Analysis of New Interactions and Flavor Conservation,''
  Nucl.\ Phys.\ B {\bf 268}, 621 (1986).

\bibitem{Colladay:1998fq}
  D.~Colladay and V.~A.~Kostelecky,
  ``Lorentz violating extension of the standard model,''
  Phys.\ Rev.\ D {\bf 58}, 116002 (1998)
  [hep-ph/9809521].\\
  V.~A.~Kostelecky and M.~Mewes,
  ``Signals for Lorentz violation in electrodynamics,''
  Phys.\ Rev.\ D {\bf 66}, 056005 (2002)
  [hep-ph/0205211].


\bibitem{WeinbergQFT}
S.~Weinberg, The Quantum Theory of Fields. Vol. 1: Foundations (Cambridge University Press, Cambridge, 1995)

\bibitem{Altschul:2003ce}
  B.~Altschul,
  ``Failure of gauge invariance in the nonperturbative formulation of massless Lorentz violating QED,''
  Phys.\ Rev.\ D {\bf 69}, 125009 (2004)
  [hep-th/0311200].
\bibitem{Altschul:2004gs}
  B.~Altschul,
  ``Gauge invariance and the Pauli-Villars regulator in Lorentz- and CPT-violating electrodynamics,''
  Phys.\ Rev.\ D {\bf 70}, 101701 (2004)
  [hep-th/0407172].
\bibitem{Altschul:2005vc}
  B.~Altschul,
  ``Radiatively induced Lorentz-violating photon masses,''
  Phys.\ Rev.\ D {\bf 73}, 036005 (2006)
  [hep-th/0512090].
\bibitem{Jackiw:1999yp}
  R.~Jackiw and V.~A.~Kostelecky,
  ``Radiatively induced Lorentz and CPT violation in electrodynamics,''
  Phys.\ Rev.\ Lett.\  {\bf 82}, 3572 (1999)
  [hep-ph/9901358].

\bibitem{PerezVictoria:2001ej}
  M.~Perez-Victoria,
  ``Physical (ir)relevance of ambiguities to Lorentz and CPT violation in QED,''
  JHEP {\bf 0104}, 032 (2001)
  [hep-th/0102021].

\bibitem{Jackiw:1999qq}
  R.~Jackiw,
  ``When radiative corrections are finite but undetermined,''
  Int.\ J.\ Mod.\ Phys.\ B {\bf 14}, 2011 (2000)
  [hep-th/9903044].

\bibitem{emergent}
   J.~Ambjorn, R.~Janik, W.~Westra and S.~Zohren,
  ``The emergence of background geometry from quantum fluctuations,''
  Phys.\ Lett.\  B {\bf 641}, 94 (2006)
  [arXiv:gr-qc/0607013].\\
  J.~Ambjorn, J.~Jurkiewicz and R.~Loll,
  ``Quantum gravity, or the art of building spacetime,''
  arXiv:hep-th/0604212.\\
  J.~Ambjorn, J.~Jurkiewicz and R.~Loll,
  ``Emergence of a 4D world from causal quantum gravity,''
  Phys.\ Rev.\ Lett.\  {\bf 93}, 131301 (2004)
  [arXiv:hep-th/0404156].\\
  Z.~C.~Gu and X.~G.~Wen,
  ``Emergence of helicity +/- 2 modes (gravitons) from qbit models,''
  arXiv:0907.1203 [gr-qc].\\
   M.~Levin and X.~G.~Wen,
  ``Colloquium: Photons and electrons as emergent phenomena,''
  Rev.\ Mod.\ Phys.\  {\bf 77}, 871 (2005).\\
 M.~Levin and X.~G.~Wen,
  ``Quantum ether: Photons and electrons from a rotor model,''
  Phys.\ Rev.\  B {\bf 73} (2006) 035122
  [arXiv:hep-th/0507118].\\
    X.~G.~Wen,
  ``Artificial light and quantum order in systems of screened dipoles,''
  Phys.\ Rev.\  B {\bf 68}, 115413 (2003)
  [arXiv:cond-mat/0210040].\
  S.~S.~Lee,
  ``Emergence of gravity from interacting simplices,''
  Int.\ J.\ Mod.\ Phys.\  A {\bf 24}, 4271 (2009)
  [arXiv:gr-qc/0609107].
 C.~Xu,
  ``Algebraic liquid phase with soft graviton excitations,''
arXiv:cond-mat/0602443.\\
  N.~Seiberg,
  ``Emergent spacetime,''
  arXiv:hep-th/0601234.\\
  L.~Sindoni, F.~Girelli and S.~Liberati,
  ``Emergent gravitational dynamics in Bose-Einstein condensates,''
  arXiv:0909.5391 [gr-qc].\\
  S.~Liberati, F.~Girelli and L.~Sindoni,
  ``Analogue Models for Emergent Gravity,''
  arXiv:0909.3834 [gr-qc].\\
  S.~Weinfurtner, M.~Visser, P.~Jain and C.~W.~Gardiner,
  ``On the phenomenon of emergent spacetimes: An instruction guide for
  experimental cosmology,''
  PoS {\bf QG-PH}, 044 (2007)
  [arXiv:0804.1346 [gr-qc]].\\
  G.~E.~Volovik,
  ``Emergent physics: Fermi point scenario,''
  Phil.\ Trans.\ Roy.\ Soc.\ Lond.\  A {\bf 366}, 2935 (2008)
  [arXiv:0801.0724 [gr-qc]].\\
  F.~R.~Klinkhamer and G.~E.~Volovik,
  ``Coexisting vacua and effective gravity,''
  Phys.\ Lett.\  A {\bf 347}, 8 (2005)
  [arXiv:gr-qc/0503090].\\
  G.~E.~Volovik,
  ``Superfluid analogies of cosmological phenomena,''
  Phys.\ Rept.\  {\bf 351}, 195 (2001)
  [arXiv:gr-qc/0005091].\\
  T.~Konopka, F.~Markopoulou and S.~Severini,
  ``Quantum Graphity: a model of emergent locality,''
  Phys.\ Rev.\  D {\bf 77}, 104029 (2008)
  [arXiv:0801.0861 [hep-th]].\\
  O.~Dreyer,
  ``Emergent general relativity,''
  arXiv:gr-qc/0604075.\\
K.~Tate and M.~Visser,
  ``Modelling gravity on a hyper-cubic lattice,''
  arXiv:1206.1092 [gr-qc].\\
M.~Visser,
  ``Survey of analogue spacetimes,''
  arXiv:1206.2397 [gr-qc].







 \bibitem{Donoghue:2010qf}
  J.~F.~Donoghue, M.~Anber and U.~Aydemir,
  ``Gauge non-invariance as tests of emergent gauge symmetry,''
  arXiv:1007.5049 [hep-ph].\\
M.~M.~Anber, U.~Aydemir and J.~F.~Donoghue,
  ``Breaking Diffeomorphism Invariance and Tests for the Emergence of Gravity,''
  Phys.\ Rev.\ D {\bf 81}, 084059 (2010)
  [arXiv:0911.4123 [gr-qc]].

 \bibitem{Weinberg:1980kq}
  S.~Weinberg and E.~Witten,
  ``Limits on Massless Particles,''
  Phys.\ Lett.\ B {\bf 96}, 59 (1980).

\bibitem{Loebbert:2008zz}
  F.~Loebbert,
  ``The Weinberg-Witten theorem on massless particles: An Essay,''
  Annalen Phys.\  {\bf 17}, 803 (2008).

\bibitem{Greisen:1966jv}
  K.~Greisen,
  ``End to the cosmic ray spectrum?,''
  Phys.\ Rev.\ Lett.\  {\bf 16}, 748 (1966).


\bibitem{Liang:2011sj}
  Y.~Liang and A.~Czarnecki,
  ``Photon-photon scattering: A Tutorial,''
  Can.\ J.\ Phys.\  {\bf 90}, 11 (2012)
  [arXiv:1111.6126 [hep-ph]].

\bibitem{Frey:2008ix}
  S.~Frey, L.~I.~Gurvits, Z.~Paragi and K.~E.~Gabanyi,
  ``High-resolution double morphology of the most distant known radio quasar at z=6.12,''
  arXiv:0805.0474 [astro-ph].
\bibitem{Shang:2011qm}
  Z.~Shang, M.~Brotherton, B.~JWills, D.~Wills, S.~Cales, D.~ADale, R.~F.~Green and J.~Runnoe {\it et al.},
  ``The New Generation Atlas of Quasar Spectral Energy Distributions from Radio to X-rays,''
  Astrophys.\ J.\ Suppl.\  {\bf 196}, 2 (2011)
  [arXiv:1107.1855 [astro-ph.CO]].

\bibitem{Particle Data Group}
	K. Nakamura et al. (Particle Data Group), JP G 37, 075021 (2010) and 2011 partial update for the 2012 edition (URL: http://pdg.lbl.gov)



\end{thebibliography}
\end{document}